# Observation of Anomalous Orbital Angular Momentum Conservation in Parametric Nonlinearity

Hai-Jun Wu[1], Bing-Shi Yu[1], Jia-Qi Jiang[1], Chun-Yu Li[1], Carmelo Rosales-Guzmán[1,2], Shi-Long Liu[3], Zhi-Han Zhu[1*], and Bao-Sen Shi[1,4*]

[1] *Wang Da-Heng Center, HLG Key Laboratory of Quantum Control, Harbin University of Science and Technology, Harbin 150080, China*
[2] *Centro de Investigaciones en Óptica, A.C., Loma del Bosque 115, Colonia Lomas del Campestre, 37150 León, Gto., Mexico*
[3] *FemtoQ Lab, Engineering Physics Department, Polytechnique Montréal, Montréal, Québec H3T 1JK, Canada*
[4] *CAS Key Laboratory of Quantum Information, University of Science and Technology of China, Hefei, 230026, China*
* e-mail: zhuzhihan@hrbust.edu.cn, and drshi@ustc.edu.cn

Orbital angular momentum (OAM) conservation plays an important role in shaping and controlling structured light with nonlinear optics. The OAM of a beam originating from three-wave mixing should be the sum or difference of the other two inputs because no light–matter OAM exchange occurs in parametric nonlinear interactions. Here, we report anomalous OAM conservation during parametric upconversion, in which a Hermite-Gauss mode signal interacts with a specially engineered pump capable of astigmatic transformation in a crystal, resulting in Laguerre-Gaussian mode sum-frequency generation (SFG). The anomaly here refers to the fact that the pump and signal carry no net OAM, while their SFG does. We show that the lost OAM with the opposite sign that maintains OAM conservation in the system is hidden in the residual pump. This unexpected OAM selection rule improves our understanding of OAM conservation in parametric nonlinear systems and may inspire new ideas for controlling OAM states via nonlinear optics, especially in quantum applications.

Since Franken *et al*. first discussed nonlinear optics more than sixty years ago, parametric nonlinearity has been extensively studied due to its irreplaceable potential in controlling the temporal frequency (or longitudinal mode) of light [1]. The term 'parametric' refers to the excited nonlinear polarization being an instantaneous virtual level, with the light-matter interaction not changing the quantum state of the medium [2]. This indicates that energy and momentum are conserved in light fields and, crucially, that the conservation law governs frequency conversion and the associated phase-matching conditions. With the increased understanding of orbital angular momentum (OAM) and structured light over the last three decades [3-5], shaping the spatial structure of light with nonlinear optics has gradually become a fascinating subject in the research community [6-8]. In addition to phase matching and polarization dependence, the effect of momentum conservation on nonlinear interactions also includes crucial contributions from OAM conservation, which has a nontrivial impact on the spatial structures of interacting beams. Thus, soon after Allen *et al*. discovered optical OAM [3], several scientists investigated the second-harmonic generation of Laguerre-Gaussian (LG) modes, noting that harmonic beams carry twice the topological charge (representing the OAM per photon) of the input. This pioneering observation provided straightforward insight into OAM conservation in parametric nonlinear systems and, more importantly, led to the development of a nonlinear optic paradigm of how OAM transfers between light [9]. The OAM of a new beam originating from a parametric nonlinear system is determined by the topological charges of the input beams. For instance, the OAM of a newly generated wave in a parametric up-/down-conversion system based on three-wave mixing is determined by the sum/difference of the OAM of the other two inputs.

Thereafter, the transfer of OAM among interacting waves (or the OAM selection rule) was widely studied in nearly all known parametric processes, from photon-level quantum interactions to ultrafast and intense-field regions, and OAM transfer between light and matter waves has also been considered [10-19]. The scope of relevant studies has recently been extended from solo OAM degree of freedom to encompass spin-orbit coupling and spatiotemporal vortices [20-27]. Remarkably, the results all follow the rooted paradigm proposed by the pioneering work. In this work, we report an unexpected *anomalous* OAM conservation in a second-order parametric nonlinear system, in which Hermite-Gaussian (HG) signals are upconverted to the corresponding LG modes by a specially engineered pump beam capable of astigmatic transformation. Compared with the current paradigm, the *anomaly* here occurs in the OAM selection rule of the three interacting waves, i.e., the pump and signal have both no net OAM during the interaction, but their upconversion becomes an LG mode carrying OAM. This unexpected result extends our current understanding of OAM conservation and may provide new insight into the nonlinear control of OAM states.

*Concept & Principle.* — To understand the principle of nonlinear astigmatic transformation (AT), we first review how to convert a correlated pair of Hermite-Laguerre-Gauss (HLG) modes, denoted as $LG_{\ell,p}$ and $HG_{m,n}$, on the same modal sphere with the order $N = 2p + |\ell| = m + n$. The crucial mathematical relation that allows this modal conversion is that $LG_{\ell,p}$ and a diagonally placed $HG_{m,n}^{45°}$ can both be represented using the superposition states of all *N*-order HG modes on the same sphere [28-31]. For simplicity, taking the 2-order HLG modes as an example, we have

$$\mathrm{HG}_{1,0}^{45°} = \sqrt{1/2}\,(\mathrm{HG}_{1,0} + \mathrm{HG}_{0,1}), \quad (1)$$
$$\mathrm{LG}_{1,0} = \sqrt{1/2}\,(\mathrm{HG}_{1,0} + i\mathrm{HG}_{0,1}). \quad (2)$$

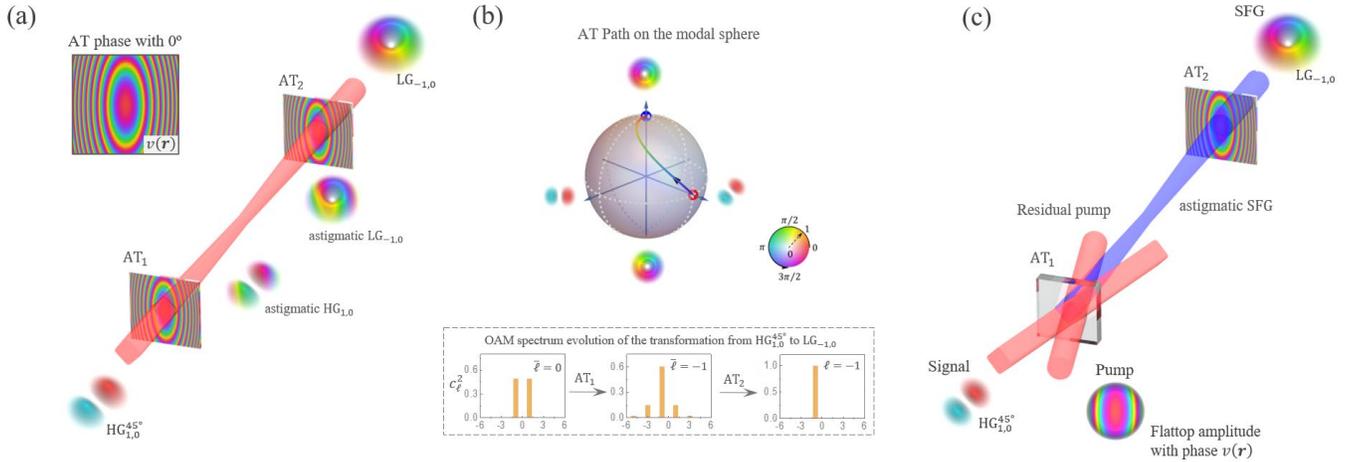

**Figure 1**. Principle of the true zero-order astigmatic transformation during HG-to-LG conversion, with (a) and (c) showing schemes for linear and nonlinear optical systems, respectively, and (b) showing the AT path on the modal sphere and the associated OAM spectrum evolution.

The conversion from Eq. (1) to Eq. (2) requires introducing a $\pm \pi/2$ intramodal phase between the two components. This unitary transformation can be achieved by exploiting the axial separability of the Gouy phase in Cartesian coordinates, which is represented as [29]

$$(N+1)\phi = (m+1/2)\phi_x + (n+1/2)\phi_y, \quad (3)$$

where $\phi$ and $\phi_{x,y}$ denote the Gouy phase and its axial components, respectively; see the *Supplementary Materials* (SM) for details. This approach mimics the rotational quarter-wave plate used in polarization control. However, in contrast to birefringence transformations, which support only unitary transformations on the Poincare sphere, ATs can be applied to control arbitrary HLG modes on higher-order modal spheres [30,31]. Notably, the AT operation realized by a cylindrical lens can provide only pseudo zero-order operations because Gouy phase accumulation inevitably occurs on the unfocused axis. True zero-order AT operations can be realized by using fractional Fourier transformations based on phase-only digital spatial light modulation [32]. Figure 1(a) shows the principle of the true zero-order AT convertor, which includes two cascading phase masks $v(\boldsymbol{r})$, as well as the beam evolution of an example mode during the conversion; see the SM for more details. When a diagonal $\mathrm{HG}_{1,0}$ mode passes through the AT convertor, the Gouy phases accumulated in the $x$ and $y$ planes are $\pi/4$ and $3\pi/4$, respectively. Thus, we have $\phi_y - \phi_x = \pi/2$, and consequently, the output beam can be converted into $\mathrm{LG}_{1,0}$, and vice versa.

In addition to modal conversion, a crucial concern in connection with OAM conservation is how the OAM transfers between the light and phase masks. In polarization control, the phase retardation and associated spin transfer gradually accumulates during birefringence propagation. However, the AT operation is entirely different, as the OAM transfer is completed in the phase mask at the HG port [29]. This conclusion can be convincingly illustrated by examining the change in the OAM spectrum, i.e., by decomposing the astigmatic beam as a set of LG modes $\sum c_{\ell,p}\mathrm{LG}_{\ell,p}$. For instance, Eq. (1) becomes $\mathrm{HG}_{1,0}^{45°}(\boldsymbol{r})e^{iv(\boldsymbol{r})}$ as the beam passes through the first mask (AT$_1$), as shown in Figure 1(b), and the corresponding OAM spectrum is broader than the original spectrum. More importantly, the spectrum is asymmetric with respect to $\ell = 0$, which indicates that the astigmatic beam $\mathrm{HG}_{1,0}^{45°}(\boldsymbol{r})e^{iv(\boldsymbol{r})}$ carries net OAM, with an average OAM of $-1\hbar$ per photon. As the beam propagates near the second mask (AT$_2$), the intramodal phases between successive LG components are modulated by the Gouy phase [33,34]. As a result, the beam structure is reshaped into an astigmatic $\mathrm{LG}_{1,0}$ mode. Then, the second mask recovers the wavefront of the astigmatic beam as a standard $\mathrm{LG}_{1,0}$ mode. This process is equivalent to compressing the OAM spectrum, and the topological charge becomes a single value, namely, $\ell = -1$.

On the basis of the above explanation, Figure 1(c) shows a schematic of a nonlinear AT modal conversion based on sum-frequency generation (SFG). Compared with the linear system, the first phase mask for performing AT$_1$ is replaced by a nonlinear crystal for the upconversion operation. The pump is a specially engineered beam that has a super Gaussian (or flattop) amplitude that carries the AT wavefront $v(\boldsymbol{r})$, which performs both frequency conversion and AT operations on the signal. In this astigmatic upconversion process, remarkably, the OAM conservation law appears to encounter a problem. Specifically, although the pump and signal both carry no net OAM (denoted as $\ell_p = \ell_s = 0$), the SFG originating from the crystal does carry OAM ($\ell_{up} \neq 0$). This finding is especially curious for the second-order all-optical nonlinear system because the nonlinear polarization in this system should be virtual and no OAM should be exchanged between the light and the crystal. In the following section, we conduct several experiments to reveal the *anomalous* OAM conservation underlying this nonlinear AT.

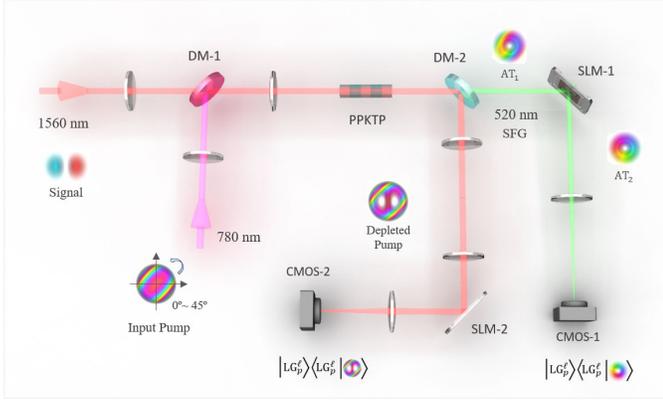

**Figure 2.** Schematic representation of the experimental setup, where the key components are the dichroic mirror (DM), spatial light modulator (SLM), camera (CMOS), and periodically poled KTiOPO4 (PPKTP) crystal.

*Experimental Results.* — Figure 2 shows a schematic setup of the nonlinear AT experiment, where we used a degenerated SFG ( signal$_{1560nm}$ + pump$_{780nm}$ → SFG$_{520nm}$) with type-0 phase matching to build the nonlinear optics platform. A pulsed (1ns 100 KHz) 1560 nm laser and its frequency doubling were used as the initial source to prepare the signal and pump. The signal beam was shaped into the desired HG mode, and the pump beam was designed with a special spatial structure — a flattop amplitude carrying the AT wavefront $v(\boldsymbol{r})$ (see the SM). The prepared HG signal was first combined with the AT pump at a dichroic mirror (DM-1) and then relayed to a 15 mm-long quasi-phase matching crystal (PPKTP) via an imaging lens. The SFG from the crystal inherits the complex amplitude of the signal and the wavefront $v(\boldsymbol{r})$ from the pump. At the output of the crystal, a short-pass filter (DM-2) separates the SFG at 520 nm from the mixing waves. This signal is then characterized by a spatial spectrum analyzer composed of a 10-bit spatial light modulator (SLM-1) and a CMOS camera [35,36]. In particular, by switching the hologram type loaded on SLM-1, i.e., a solo phase mask for LG projections or a combination mask including $v(\boldsymbol{r})$, we can measure the OAM spectrum of the SFG before and after the SFG passes through the second AT phase mask. In the reflecting port of DM-2, the residual pump is relayed to another spatial spectrum analyzer to measure the OAM. In addition, we use a complex amplitude profiler based on spatial Stokes tomography for in situ observations of the full spatial structure of the three beams (see the SM).

In the experiment, for convenience, we used a rotational AT pump to interact with horizontal HG signals instead of the rotated signal scheme shown in Figure 1. Without loss of generality, two low-order HG modes, HG$_{1,0}$ and HG$_{2,0}$, were chosen as the example signals. The angle of the AT pump with respect to the horizontal plane was set to 22.5° and 45°, to convert HG$_{1,0}$ and HG$_{2,0}$ to intermediate HLG modes and LG$_{1,0}$ and LG$_{2,0}$ modes on the modal sphere, respectively. We first consider a small signal scenario in which the amplitude of the AT pump is assumed to be constant. Figure 3 shows the measured complex amplitude and OAM spectra of the input HG signal and associated HLG and LG SFG, as well as their theoretical references. The results, including both measured OAM spectra and observed complex amplitudes, of the original HG signals, intermediate HLG and final LG SFG are consistent with the theorical predictions. Notice that the one-dimensional OAM spectrum ($c_\ell^2$) is contracted from the two-dimensional LG spectrum ($c_{\ell,p}^2$) provided in the SM. Moreover, the amount of net OAM carried by the two SFG beams, denoted as $\ell_{up}\hbar$, gradually increased to $-\hbar$ and $-2\hbar$ per photon as the angle of the AT pump increased from 22.5° to 45°. More importantly, the net OAM inflows to the SFG beams occurred during the first nonlinear AT operation in the crystal, while the second linear AT operation implemented by SLM-1 removed the AT wavefront of the beam while compressing the OAM spectra. This result, however, poses a significant challenge to the OAM conservation law in parametric nonlinear systems; namely, since the two inputs in the three-wave interaction carry no net OAM, what is the origin of the OAM inflows in the third wave during the interaction?

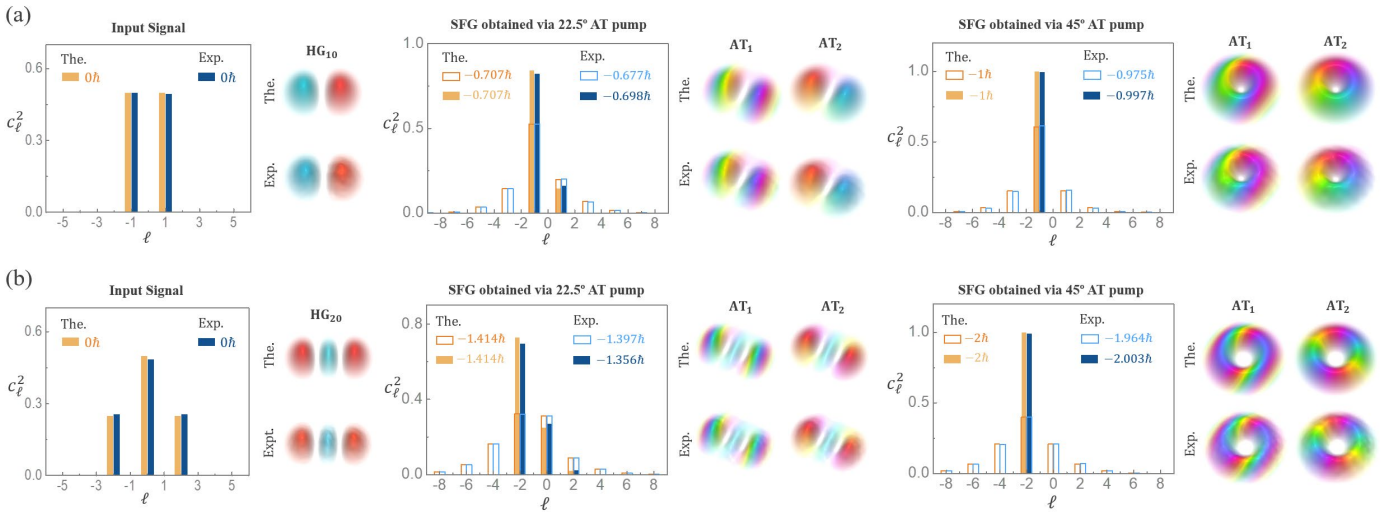

**Figure 3.** Simulated and measured OAM spectra of the input HG signal and associated SFG outputs, as well as their complex amplitudes, where (a) and (b) show the results corresponding to using weak HG$_{1,0}$ and HG$_{2,0}$ modes as signals, respectively. The patterns in AT$_1$ (AT$_2$) demonstrate the complex amplitudes of the SFG signal before (after) the second astigmatic transformation, and the unfilled (filled) histograms represent the corresponding uncompressed (compressed) OAM spectra.

The SFG is an all-optical parametric nonlinearity that has no light-matter OAM exchange. Moreover, previous works have shown that the complex amplitude of structured signal can be maintained during flattop-beam pumped upconversion [37], implying that no OAM changes occur in the unconverted signal. Therefore, the only possible hideout of the undiscovered OAM that can conserve the OAM conservation in the system is the residual pump beam. The interactions with the HG modes reshape the amplitude structure of the residual pump, which may lead to this beam carrying net OAM. To verify this hypothesis, we consider a pump depletion case, i.e., a weak pump interacting with strong signals, to clarify the change in the amplitude structure of the residual pump. From a theoretical perspective, we assumed that the amplitude of the residual pump was completely depleted by the peak amplitude of the signal; see the SM for details. To reproduce this assumption experimentally, by tuning the average power of the pump (1 mW) and signal (~25/30mw for HG10/HG20) beams, we ensured that the observed beam profiles of the residual pump were as close as possible to that of the theoretical reference.

Figure 4 shows the measured OAM spectra of the AT pump before and after the depletion interaction, as well as the corresponding complex amplitudes $a_p(\boldsymbol{r})$ and $a_{p'}(\boldsymbol{r})$ observed during the same exposure time. For all cases, the original pump has a perfect flattop amplitude and a symmetric OAM spectrum with respect to the $\ell = 0$ axis, confirming that the beam carries no OAM ($\ell_p = 0$). After the interaction, however, the pump depletion resulted in a reduced amplitude in the interaction regions, exhibiting holes with the same shapes as the signal profiles, i.e., the $HG_{1,0}$ and $HG_{2,0}$ modes. This depletion caused the OAM spectra of the residual pump to become asymmetric with respect to $\ell = 0$, which is a feature of a beam carrying net OAM and thus qualitatively confirms the hypothesis.

To quantitatively explore the OAM selection rule, we focus on OAM conservation in the system. This requires the total amount of OAM inflow in the residual pump ($\ell_{p'}$) should be equal to that in the generated SFG but with the opposite sign, which can be expressed as

$$\ell_{up}\bar{n}_{up} = -\ell_{p'}\bar{n}_{p'}. \qquad (4)$$

Here, $\bar{n}_{up}$ and $\bar{n}_{p'}$ denote the average number of photons contained in the SFG and residual pump beams, respectively, and their ratio can be calculated with the following equation (see the SM for details):

$$\frac{\bar{n}_{up}}{\bar{n}_{p'}} = \frac{\iint [a_p^2(\boldsymbol{r}) - a_{p'}^2(\boldsymbol{r})]d\boldsymbol{r}}{\iint a_{p'}^2(\boldsymbol{r})\,d\boldsymbol{r}}, \qquad (5)$$

where $a_p^2(\boldsymbol{r})$ and $a_{p'}^2(\boldsymbol{r})$ have the same peak power in plane $\boldsymbol{r}$, corresponding to patterns recorded in experiments by using a camera with the same exposure time. The term $\iint [a_p^2(\boldsymbol{r}) - a_{p'}^2(\boldsymbol{r})]d\boldsymbol{r}$ describes the variation in the pump power, which is proportional to the number of lost photons, and these lost photons are converted into SFG signals (i.e., $\bar{n}_{up}$). By assuming $\iint a_p^2(\boldsymbol{r})\,d\boldsymbol{r} = 1$, the power of the residual pump $\iint a_{p'}^2(\boldsymbol{r})\,d\boldsymbol{r}$ in the two cases, i.e., using the $HG_{1,0}$ and $HG_{2,0}$ modes as signals, are equal to approximately 0.753 and 0.812, respectively, and the corresponding ratios are 0.328 and 0.232. We can thus use Eq. (4) to obtain the expected average OAM carried by the 45° (22.5°) residual pump, i.e., $\ell_{p'}\hbar$ per photon, which is equal to approximately $0.328\hbar$ ($0.232\hbar$) and $0.464\hbar$ ($0.328\hbar$), respectively. These theoretical predictions are consistent with the results calculated according to the measured OAM spectra. The deviation between the theory and experiments originates from the finite dimensional projections and imperfect beam structure.

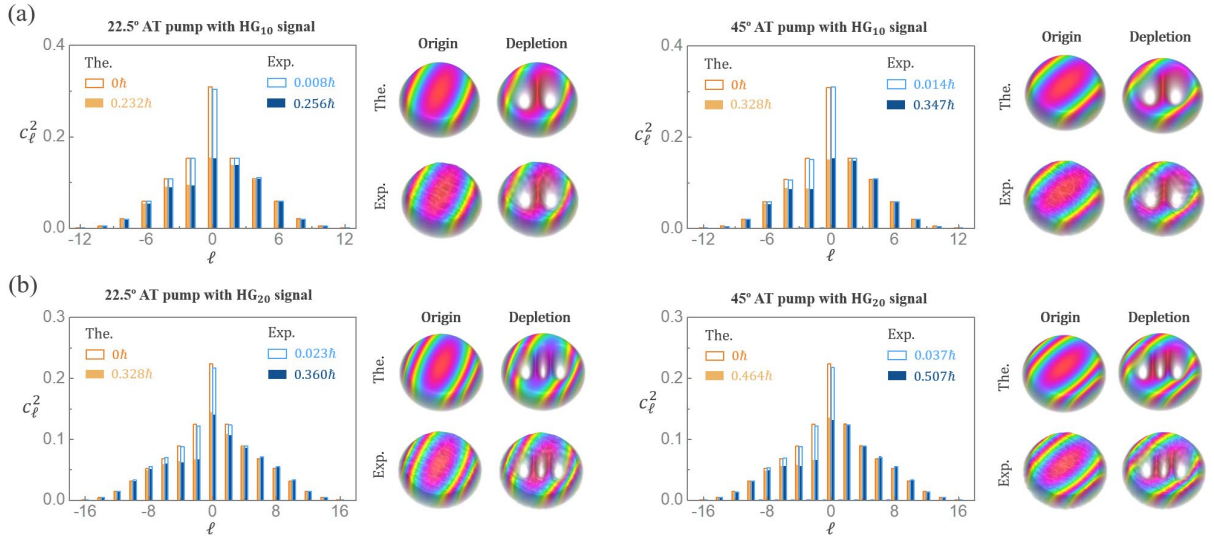

**Figure 4.** Simulated and measured OAM spectra of the original (unfilled histograms) and residual (filled histograms) pumps, as well as their complex amplitudes, where (a) and (b) show the results corresponding to using intense $HG_{1,0}$ and $HG_{2,0}$ modes as signals, respectively.

*Conclusion.* — We report experimentally anomalous OAM conservation in nonlinear AT operations, in which HG signals were upconverted into LG or intermediate HLG modes according to the relative angle of the AT pump. This demonstration provides a useful nonlinear technique for shaping structured light and, more importantly, reveals an unexpected OAM selection rule, namely, that the pump and signal both carry no net OAM, but their SFG does carry OAM, which contradicts the current paradigm. Our results show that the residual pump carries the same OAM as the SFG but with an opposite chirality, thus maintaining OAM conservation in the system. These findings provide deeper insight into OAM conservation in parametric nonlinear systems and indicate that similar phenomena should be explored, such as analogs in parametric amplification and four-wave mixing [11,15,24]. Moreover, many exciting results in quantum nonlinear systems should be investigated. More specifically, the interaction can be regarded as a nonlinear OAM sorter that separates the pump beam into two frequency bins carrying opposite net OAM. Thus, perhaps it can convert a polarization entanglement into a hybrid one including both frequency and OAM degree of freedoms. Besides, the full even-valued OAM components in the AT pump indicates astigmatic wavefront may be used for shaping the high-dimension state emitting from a spontaneous parametric down-conversion [12,38].

**Acknowledgments**

This work was supported by the National Natural Science Foundation of China (Grant Nos. 62075050, 11934013, and 61975047) and the High-Level Talents Project of Heilongjiang Province (Grant No. 2020GSP12).